\documentclass[%
 reprint,
superscriptaddress,
 amsmath,amssymb,
 aps,
pra,
]{revtex4-2}

\usepackage{graphicx}
\usepackage{dcolumn}
\usepackage{bm}
\usepackage{xfrac}
\usepackage{xcolor}

\begin{document}
\newcommand{\edit}[1]{\textcolor{black}{#1}}
\newcommand{\StanfordAP}{Department of Applied Physics, Stanford University, Stanford, CA 94305, USA}
\newcommand{\StanfordEE}{Department of Electrical Engineering, Stanford University, Stanford, CA 94305, USA}
\newcommand{\StanfordPhysics}{
Department of Physics, Stanford University, Stanford, CA 94305, USA}
\newcommand{\PULSE}{
Stanford PULSE Institute, SLAC National Accelerator Laboratory, Menlo Park, CA 94025, USA}
\newcommand{\SLAC}{Linac Coherent Light Source, SLAC National Accelerator Laboratory, Menlo Park, CA 94025, USA}
\newcommand{\Edinburgh}{EaStCHEM School of Chemistry, University of Edinburgh, Edinburgh EH9 3FJ, United Kingdom}
\newcommand{\Michigan}{Department of Chemistry, University of Michigan, Ann Arbor, MI, 48109, USA}
\newcommand{\SRI}{Stanford Research Institute, Menlo Park, CA, 94025, USA}
\newcommand{\Brown}{Department of Chemistry, Brown University, Providence, RI, 02912, USA}
\newcommand{\IIT}{Department of Physics, Indian Institute of Technology Bombay, Powai, Mumbai, 400076, India}
\newcommand{\Southampton}{School of Chemistry, University of Southampton, Highfield, Southampton SO17 1BJ, UK}
\newcommand{\OxfordAK}{Physical and Theoretical Chemistry Laboratory, Department of Chemistry, University of Oxford, South Parks Road, OX1 3QX Oxford, UK}
\newcommand{\KTH}{Department of Chemistry, KTH Royal Institute of Technology, SE-10044 Stockholm, Sweden}
\newcommand{\UoB}{School of Chemistry, University of Birmingham,  Birmingham B15 2TT, UK}
\newcommand{\EuXFEL}{European XFEL, Holzkoppel 4, 22869 Schenefeld, Germany}
\newcommand{\DESY}{Deutsches Elektronen-Synchrotron DESY, Hamburg, Germany}
\newcommand{\Davis}{Department of Chemistry, University of California, Davis, One Shields Avenue, Davis CA 95616}

\title{Imaging valence electron rearrangement in a chemical reaction using hard X-ray scattering}

\author{Ian Gabalski}
\email{igabalsk@stanford.edu (Experiment)}
\affiliation{\PULSE}
\affiliation{\StanfordAP}
\author{Alice Green}
\affiliation{\PULSE}
\affiliation{\EuXFEL}
\affiliation{\Edinburgh}
\author{Philipp Lenzen}
\affiliation{\PULSE}
\author{Felix Allum}
\affiliation{\PULSE}
\affiliation{\SLAC}
\affiliation{\DESY}
\author{Matthew Bain}
\affiliation{\SLAC}
\author{Surjendu Bhattacharyya}
\affiliation{\PULSE}
\author{Mathew A. Britton}
\affiliation{\SLAC}
\author{Elio G. Champenois}
\affiliation{\PULSE}
\author{Xinxin Cheng}
\affiliation{\SLAC}
\author{James P. Cryan}
\affiliation{\PULSE}
\affiliation{\SLAC}
\author{Taran Driver}
\affiliation{\PULSE}
\affiliation{\SLAC}
\author{Ruaridh Forbes}
\affiliation{\Davis}
\affiliation{\SLAC}
\author{Douglas Garratt}
\affiliation{\SLAC}
\author{Aaron M. Ghrist}
\affiliation{\PULSE}
\affiliation{\StanfordAP}
\author{Martin Gra{\ss}l}
\affiliation{\PULSE}
\affiliation{\SLAC}
\author{Matthias F. Kling}
\affiliation{\SLAC}
\author{Kirk A. Larsen}
\affiliation{\SLAC}
\author{Mengning Liang}
\affiliation{\SLAC}
\author{Ming-Fu Lin}
\affiliation{\SLAC}
\author{Yusong Liu}
\affiliation{\SLAC}
\author{Michael P. Minitti}
\affiliation{\SLAC}
\author{Silke Nelson}
\affiliation{\SLAC}
\author{Joseph S. Robinson}
\affiliation{\SLAC}
\author{Philip H. Bucksbaum}
\affiliation{\PULSE}
\affiliation{\StanfordAP}
\affiliation{\StanfordPhysics}
\author{Thomas J. A. Wolf}
\affiliation{\PULSE}
\author{Nanna H. List}
\email{nalist@kth.se; n.h.list@bham.ac.uk (Theory)}
\affiliation{\KTH}
\affiliation{\UoB}
\author{James M. Glownia}
\affiliation{\SLAC}

\date{\today}

\begin{abstract}
We have observed the signatures of valence electron rearrangement in photoexcited ammonia using ultrafast hard X-ray scattering. Time-resolved X-ray scattering is a powerful tool for imaging structural dynamics in molecules because of the strong scattering from the core electrons localized near each nucleus. Such core-electron contributions generally dominate the differential scattering signal, masking any signatures of rearrangement in the chemically important valence electrons. Ammonia represents an exception to the typically high core-to-valence electron ratio. We measured 9.8 keV X-ray scattering from gas-phase deuterated ammonia following photoexcitation via a 200 nm pump pulse to the 3s Rydberg state. We observed changes in the recorded scattering patterns due to the initial photoexcitation and subsequent deuterium dissociation. \textit{Ab initio} calculations confirm that the observed signal is sensitive to the rearrangement of the single photoexcited valence electron as well as the interplay between adiabatic and nonadiabatic dissociation channels. The use of ultrafast hard X-ray scattering to image the structural rearrangement of single valence electrons constitutes an important advance in tracking valence electronic structure in photoexcited atoms and molecules.
\end{abstract}

\maketitle


The spatial distribution of valence electron density is key in determining chemical bonding, molecular structure, and chemical reactivity. 
The presence or absence of electron density between atoms in a molecule determines whether a chemical bond will be formed \cite{pauling_nature_1960,bag_attochemistry_2021}, while the electron cloud's spatial symmetry and localization determines their interaction with light, for example in single-photon excitation \cite{bunker_molecular_2006,barone_computational_2021} and strong-field multi-photon and tunnel ionization \cite{cheng_momentum-resolved_2020,howard_strong-field_2021,howard_filming_2023}.
Chemical reaction pathways are determined by nuclear motion within a potential that is generated by the adiabatic adjustment of the valence electron density to the changing nuclear positions. 
In the case of ultrafast dynamics through conical intersections, nuclear motion can also drive nonadiabatic changes in the valence electron density \cite{worth_beyond_2004}. This interplay of valence electronic and nuclear degrees of freedom drives some of the most important photochemical reactions in nature, e.g.~the initial processes in photosynthesis and human vision \cite{cheng_dynamics_2009,schoenlein_first_1991,polli_conical_2010}.  Yet, despite the importance of the spatial distribution of valence electron density to all aspects of chemistry, direct observation of their time-resolved dynamics during a chemical reaction has remained an elusive goal.


\begin{figure}[t]
    \centering
    \includegraphics[width=0.45\textwidth]{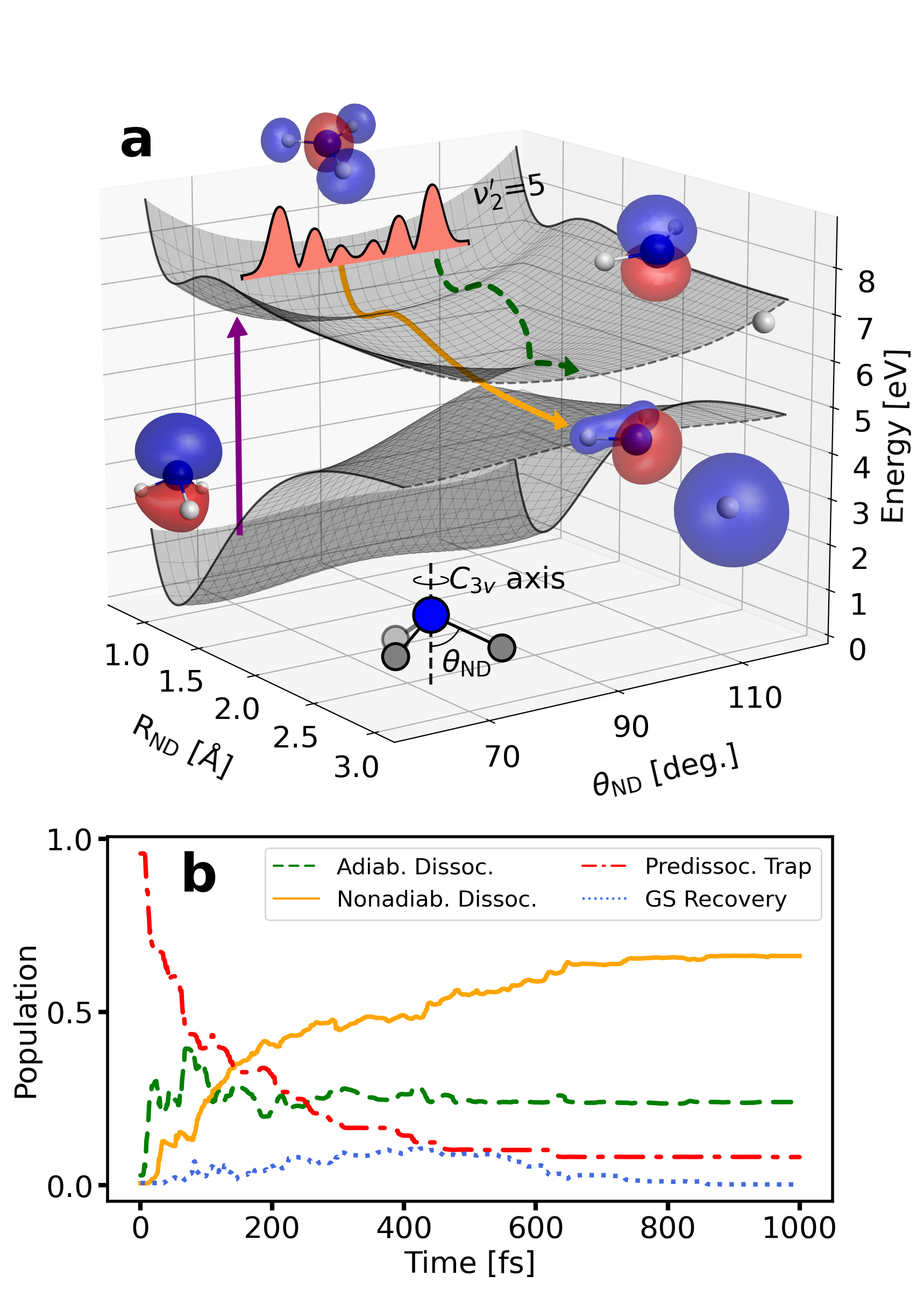}
    \caption{(a) Potential energy surfaces of ND$_3$ as a function of $\nu_2^\prime$ umbrella bending angle $\theta_\mathrm{ND}$ and N-D bond length $R_\mathrm{ND}$. Pump process (\edit{purple} solid vertical), adiabatic (green dashed), and nonadiabatic (orange solid) arrows show the reaction pathways. Representative nuclear geometries and orbital structures 
    are rendered using wxMacMolPlt \cite{bode_macmolplt_1998}. 
    Approximate $\nu_2^\prime=5$ wavefunction is illustrated in the Franck-Condon region. Inset: diagram illustrating the $\theta_\mathrm{ND}$ coordinate associated with the $\nu_2^\prime$ vibrational mode. (b) AIMS-computed time-dependent populations of the four primary reaction channels: photoexcited predissociation population \edit{(red dot-dashed)}, adiabatic \edit{(green dashed)} and nonadiabatic \edit{(orange solid)} dissociation, and ground-state recovery \edit{(blue dotted)}.}
    \label{fig:pes and diagram}
\end{figure}

The experimental barriers to measuring the time-evolving valence electron structure in a chemical reaction can be attributed to the fact that most ultrafast probes lack direct sensitivity to this molecular property. Time-resolved spectroscopy probes transitions between molecular energy levels and their associated cross-sections \cite{weinacht_time-resolved_2018}, from which the spatial structures of the associated orbitals can sometimes be inferred through an analysis of their angular dependence \cite{bisgaard_time-resolved_2009,hockett_time-resolved_2011,itatani_tomographic_2004}. The element specificity of time-resolved X-ray spectroscopy can be employed to monitor local valence electron density changes at specific atomic sites \cite{mayer_following_2022,leitner_time-resolved_2018,gabalski_time-resolved_2023,wolf_probing_2017}. Although these spectroscopic probes can reveal information related to electron density, they lack the direct sensitivity to the overall spatial distribution of the valence electrons. Meanwhile, time-resolved probes with Angstrom-scale wavelengths such as megaelectronvolt ultrafast electron diffraction (MeV-UED) \cite{centurion_ultrafast_2022,yang_femtosecond_2016,shen_femtosecond_2019,wolf_photochemical_2019,champenois_femtosecond_2023} and time-resolved X-ray scattering (TRXS) \cite{stankus_advances_2020,stankus_ultrafast_2019,yong_observation_2020,gabalski_transient_2022,bucksbaum_characterizing_2020} can spatially resolve Angstrom-size structures in molecular systems. \edit{These diffractive probes nevertheless} interact with all the electrons in the system and are thus dominated by core-electron and nuclear scattering contributions in most cases. 

Several studies have observed the signatures of valence electronic rearrangement using scattering probes. Yang \textit{et al.}~\cite{yang_femtosecond_2016} and Champenois \textit{et al.}~\cite{champenois_femtosecond_2023} utilized MeV-UED to probe photoexcited pyridine and deuterated ammonia, respectively. Both studies observed inelastic scattering signatures from electron-electron correlation at low scattering angles due to valence electronic structure changes induced by the photoexcitation. They demonstrated that the MeV-UED technique is capable of separating electron and nuclear correlation signatures to different regions of scattering angle. However, the technique currently suffers from the lack of time resolution ($>$150~fs \cite{shen_femtosecond_2019}). Moreover, inelastic scattering is incoherent and, therefore, does not contain direct information about the changes in the spatial valence electron distribution from the photoexcitation \cite{yang_simultaneous_2020}.

\begin{figure*}[t]
    \centering
    \includegraphics[width=0.8\textwidth]{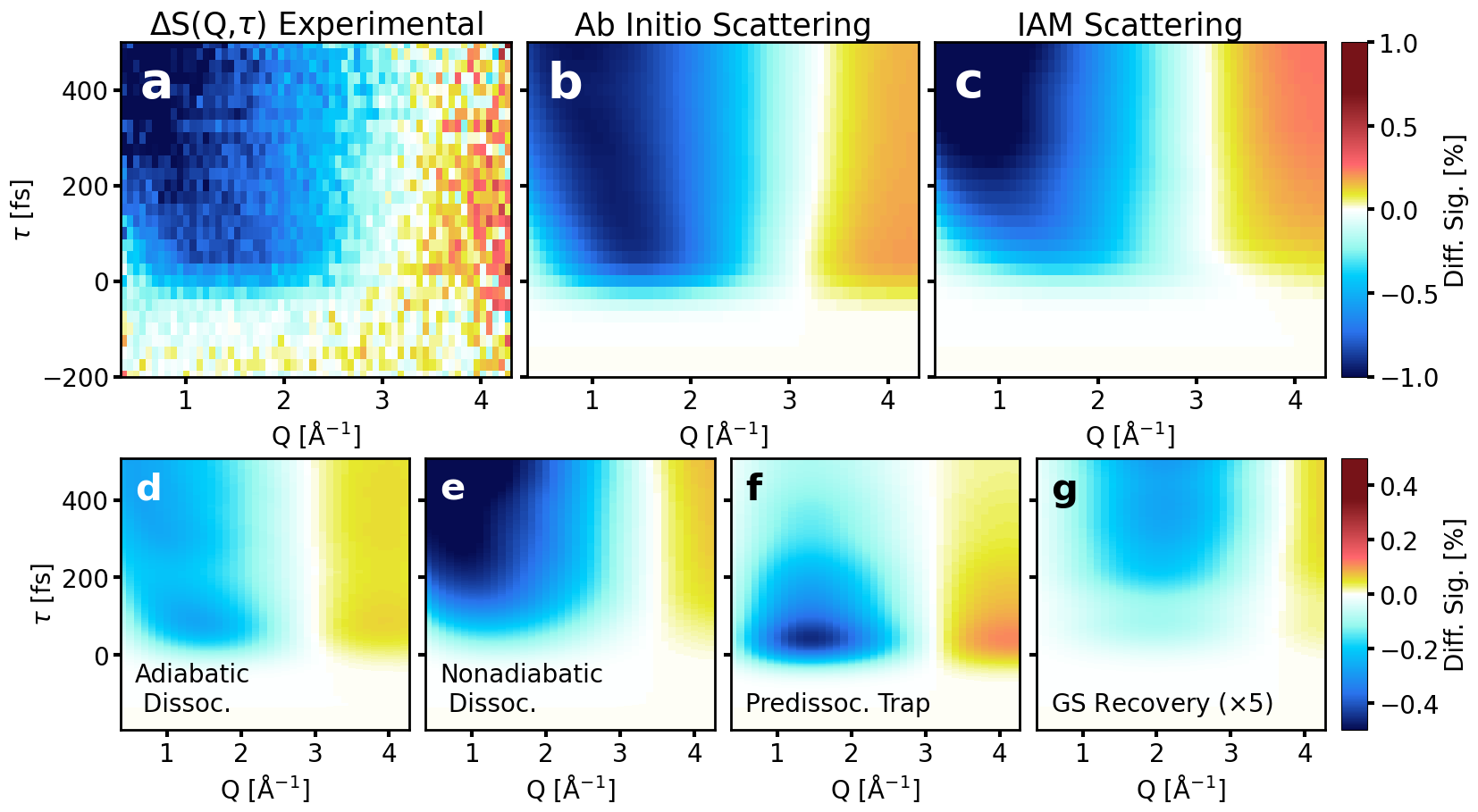}
    \caption{(a) Experimental difference signal. (b)-(c) \textit{Ab initio} and IAM total scattering difference signals, respectively, convolved in time with the experimentally determined 80~fs FWHM instrument response. (d)-(g) \textit{Ab initio} signal decomposed into four channels: adiabatic and nonadiabatic dissociation, predissociation trapping of the Rydberg state in the Franck-Condon region, and recovery of population in the bound region of the ground electronic state. 
    }
    \label{fig:difference signals}
\end{figure*}

TRXS is capable of providing superior time resolution ($\sim$10~fs or better \cite{glownia_pumpprobe_2019}) and is sensitive exclusively to the electron density distribution. Recent TRXS studies by Yong \textit{et al.}~and Stankus \textit{et al.}~reported observation of signatures from changes in the valence electron density distribution due to photoexcitation to molecular Rydberg states \cite{yong_observation_2020, stankus_ultrafast_2019}. Another study by Yong \textit{et al.} reported 
\edit{signatures} of charge transfer dynamics on picosecond timescales via TRXS \cite{yong_ultrafast_2021}. The observation of these signatures is complicated by the relative weakness of the scattering signatures from the delocalized valence electrons with respect to the signatures of the core electrons.
Thus, observation of changes in the valence electron distribution
could only be obtained within the first tens of femtoseconds before the onset of nuclear motion\edit{, or inferred indirectly from their effects on the nuclear geometry. TRXS probes}
have yet to achieve full tracking of the valence electron dynamics throughout an entire chemical reaction.

Here we report results from a joint experimental-theoretical TRXS study on the photodissociation of deuterated ammonia (ND$_3$). The lack of significant scattering contributions in this system from numerous localized core electrons allows us to exclusively measure the time-dependent evolution of the valence electron distribution throughout the entire reaction. Gas-phase ND$_3$ photoexcited by a 6.2~eV (200~nm) photon is driven to the 2$^1A$ state, a transition \edit{illustrated in Fig.~\ref{fig:pes and diagram}(a)} that is dominated by the excitation of a lone pair electron to the 3s Rydberg orbital. 
A low barrier to D$_2$N-D neutral dissociation leads to an excited-state lifetime on the order of 100~fs \cite{vaida_ultraviolet_1987-1,chatterley_timescales_2013}. Past the predissociation barrier, the excited state crosses with the ground electronic state near the D$_2$N-D stretched C$_{2v}$ geometry, yielding both adiabatic and nonadiabatic dissociation channels (green dashed and orange solid arrows in Fig.~\ref{fig:pes and diagram}(a)). \edit{Figure~\ref{fig:pes and diagram}(b) shows the time-resolved populations of the involved reaction channels computed using \textit{ab initio} multiple spawning (AIMS) \cite{ben-nun_ab_2000} simulations, as described in the End Matter.}

\edit{We performed a TRXS measurement on photoexcited ND$_3$ at the Linac Coherent Light Source (LCLS) that is described in the End Matter.} The time-resolved \edit{scattering} difference signal was computed,
$\Delta S(Q,\tau)~=~(S(Q,\tau)-S_0(Q))/S_0(Q)$
where \edit{$Q$ is the scattering momentum transfer, $\tau$ is the pump-probe delay, and }$S_0(Q)$ is the average scattering signal for the un-pumped ND$_3$ sample obtained on a subset of all FEL shots interspersed through the data collection. 
Accurate shot-to-shot estimates of the X-ray pulse energy obtained by upstream diode measurements, described in detail in the Supplemental Material \cite{noauthor_see_nodate}, were crucial for correct scattering signal normalization. The resulting difference signal $\Delta S(Q,\tau)$ in Fig.~\ref{fig:difference signals}(a) shows a prompt depletion around time zero between $Q=0.7$ and 2.6~\AA$^{-1}$ as well as a smaller enhancement above $Q=3.0$~\AA$^{-1}$. Within the first 200~fs, the depletion widens to both lower and higher $Q$ and the enhancement shrinks in $Q$ before becoming roughly static, indicating that most of the relevant dynamics occur in this time delay region. Careful analysis of this region reveals that much of the change in scattering signal is due to the evolving valence electronic distribution.

The contribution to the scattering signal due to chemical bonding can be differentiated via comparison to the equivalent signals computed with both fully \textit{ab initio} scattering calculations and the independent atom model (IAM). Both the \textit{ab initio} and IAM calculations of the expected scattering were performed using the AIMS molecular trajectories, the results of which are shown in Fig.~\ref{fig:difference signals}(b-c). The IAM signal approximates the molecule as an assembly of non-interacting atoms, ignoring any valence electron density distortions due to chemical bonding. The atoms then scatter coherently, and the modulations in the scattering can be related to the pairwise internuclear distances \cite{stankus_advances_2020,brown_intensity_2006}:
\begin{equation}
    S_{\mathrm{IAM}}(Q)=\sum_{\alpha=1}^{N_{at}}\left|f_\alpha(Q)\right|^2 + \sum_{\alpha\neq\beta}^{N_{at}}f_\alpha(Q)f_\beta(Q)\frac{\sin(Qr_{\alpha\beta})}{Qr_{\alpha\beta}}
\end{equation}
Here $r_{\alpha\beta}$ is the distance between atoms $\alpha$ and $\beta$, $f_\alpha(Q)$ is the scattering form factor of atom $\alpha$ \cite{brown_intensity_2006}, and the expression shown is for an isotropic ensemble of molecules. The fully \textit{ab initio} scattering calculation, meanwhile, takes full account of the electron density distributions involved in chemical bonding as well as both elastic and inelastic scattering \cite{parrish_ab_2019,ufimtsev_quantum_2008, ufimtsev_quantum_2009,ufimtsev_quantum_2009-1,seritan_terachem_2021}. It should be noted that the inelastic component here has negligible contribution to the total X-ray scattering signal, distinguishing it from the MeV-UED signal of the same photochemical reaction \cite{champenois_femtosecond_2023}.

The AIMS trajectories and their corresponding scattering contributions can be separated into four distinct classes based on simple criteria: adiabatic dissociation, nonadiabatic dissociation, predissociation trapping, and ground state recovery. All of the photoexcited population begins in the predissociation trap, remaining behind the predissociation barrier in the Franck--Condon region while slowly leaking out into one of the two dissociation channels. Once population has crossed the predissociation barrier at $R_\mathrm{ND}=1.33$~$\mathrm{\AA}$, it first enters the adiabatic channel leading up to the conical intersection at $R_\mathrm{ND}=2.07$~$\mathrm{\AA}$. \edit{Following passage through} the conical intersection, population splits into either the adiabatic or nonadiabatic dissociation channels, which undergo further N-D bond \edit{elongation} and leave the remaining ND$_2$ radical in either its electronic ground (ND$_2$($\tilde X$)) or lowest excited (ND$_2$($\tilde A$)) state, respectively. The ground state recovery channel is the small portion of the population that evolves along the dissociation coordinate up to the conical intersection, but then returns to the bound ground state potential well and remains bound. The decomposition of the \textit{ab initio} scattering difference signal into these four channels is shown in Figs.~\ref{fig:difference signals}(d-g). 

\begin{figure}[t]
    \centering
    \includegraphics[width=0.4\textwidth]{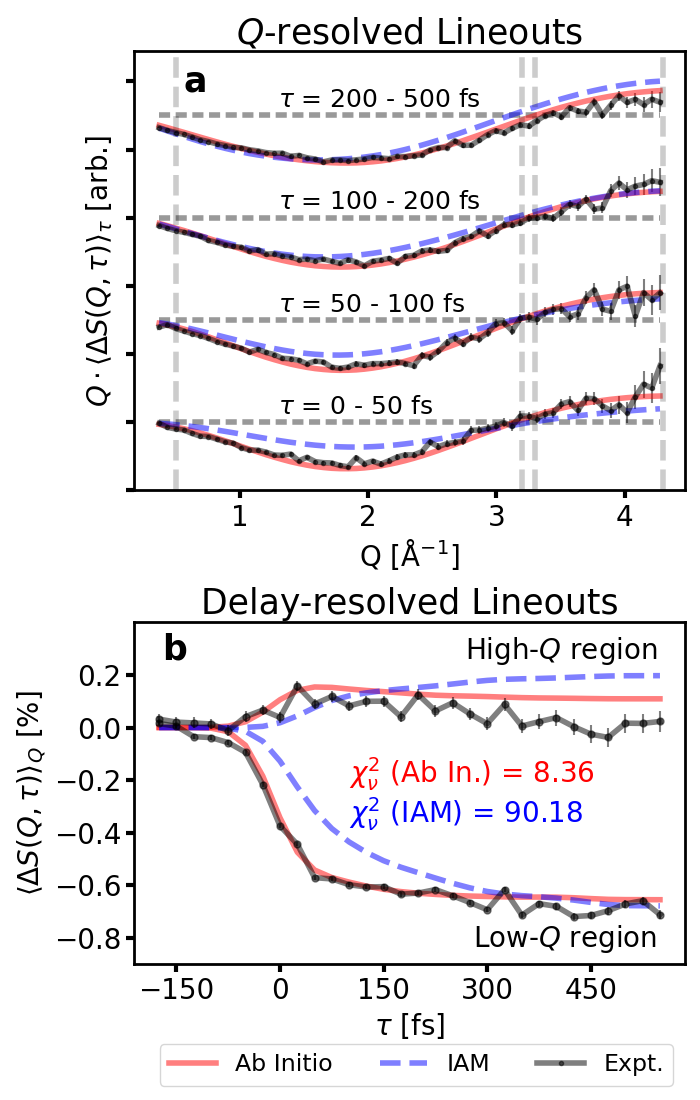}
    \caption{(a) $Q$-resolved lineouts of the experimental data (black \edit{points with $1\sigma$ error bars}), \textit{ab initio} (red \edit{solid}), and IAM (blue \edit{dashed}) difference signals for a selected set of pump-probe delays. Difference signals have been scaled by $Q$ to enhance feature visibility above $Q=3$~$\mathrm{\AA}^{-1}$. \edit{Vertical dashed gray lines indicate boundaries of the lineouts presented in the next panel.} (b) Delay-resolved average difference signals in the \edit{$0.5$~$\mathrm{\AA}^{-1}$$<Q<3.2$~$\mathrm{\AA}^{-1}$} and \edit{$3.3$~$\mathrm{\AA}^{-1}$$<Q<4.3$~$\mathrm{\AA}^{-1}$} regions bounded by the vertical dashed lines in \edit{panel }(a).}
    \label{fig:lineouts}
\end{figure}

Figure~\ref{fig:lineouts} shows the experimental data alongside both the \textit{ab initio} and IAM calculations of the signal in two ways. Figure~\ref{fig:lineouts}(a) compares the experimental difference signal to the \textit{ab initio} (red) and IAM (blue) signals within particular pump-probe delay regions. The experimental $Q$-resolved lineouts are shown in black with bootstrapping-estimated 1$\sigma$ error bars, which are too small to be distinguished from the plotted markers for all but the highest-$Q$ regions. 
\edit{Two independent least-squares fits of the data to both the \textit{ab initio} and IAM difference signals were performed with the excitation fraction as the free parameter, yielding estimated excitation fractions of 6.6\% and 8.6\%, respectively.}
The experimental data are generally more consistent with the \textit{ab initio} signal than with the respective IAM signal\edit{, even allowing for independent estimation of the excitation fraction}.

Figure~\ref{fig:lineouts}(b) further elucidates the differences between the experimental, \textit{ab initio}, and IAM signals by taking \edit{delay}-resolved lineouts of the low- and high-$Q$ regions of $\Delta S(Q,\tau)$, the bounds of which are indicated by the vertical \edit{gray} dashed lines in Fig.~\ref{fig:lineouts}(a). The lineout of the low-$Q$ depletion region between $0.5$~$\edit{\mathrm{\AA}^{-1}}$$<Q<3.2$~$\mathrm{\AA}^{-1}$ demonstrates the superiority of the fit of the \textit{ab initio} signal to the data compared to that of the IAM initially after photoexcitation. 
The faster rate of depletion in the ab initio case (and the experiment) is due to 
the \edit{prompt electronic} excitation to the 3s Rydberg orbital. In contrast, the changes in the IAM signal require the deuterons to have moved significantly, which takes a few hundred femtoseconds. The computed scattering contribution from the bound, photoexcited ND$_3$ population shown in Fig.~\ref{fig:difference signals}(f) demonstrates \edit{that} this prompt depletion below $Q=3.2$~$\mathrm{\AA}^{-1}$ is due to the n3s-photoexcited electron. The excess low-$Q$ depletion in the \textit{ab initio} and experimental scattering lasts for at least 150~fs before the motion of the deuterons captured by the IAM signal begins to sufficiently account for the scattering change.

The region of enhancement between $3.3$~\edit{\AA$^{-1}$}$<Q<4.3$~\AA$^{-1}$ also exhibits distinct signatures of the valence electron dynamics. In the experimental data, the enhancement initially rises, then begins to fall again after 100~fs. This behavior is qualitatively reproduced by the \textit{ab initio} theory, but not by the IAM signal that shows a monotonic rise. The qualitative discrepancy between the IAM signal and the experimental data demonstrates that the dominant effect we observe here is the dynamics of the \edit{photoexcited} valence electron rather than the structural dynamics of the deuteron dissociation. 

\begin{figure}[t]
    \centering
    \includegraphics[width=0.45\textwidth]{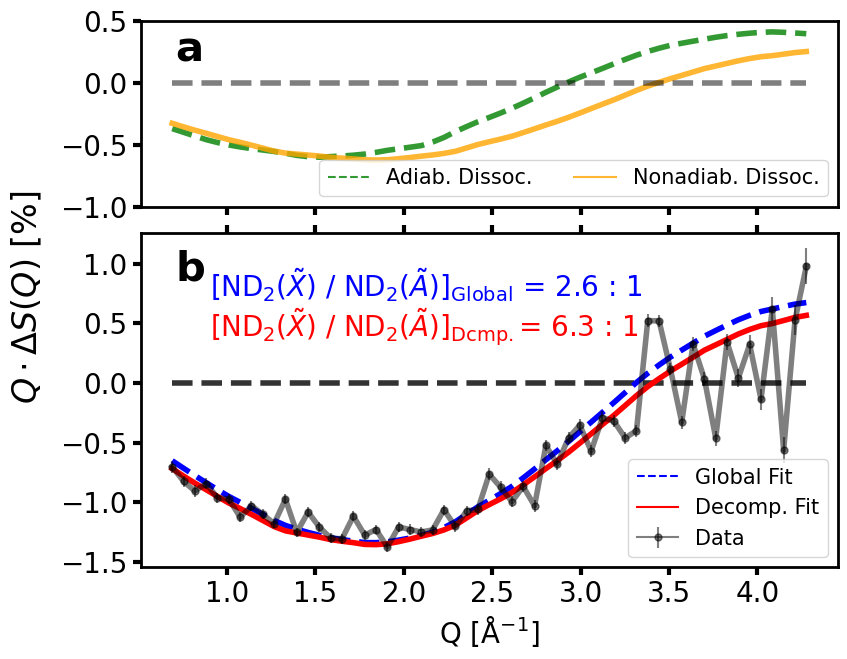}
    \caption{(a) $Q$-resolved lineouts of the long-delay ($\tau>500$~fs) \textit{ab initio} difference signal decomposed into adiabatic (green \edit{dashed}) and nonadiabatic (orange \edit{solid}) dissociation channels. (b) Long-delay data (black \edit{points with $1\sigma$ error bars}) compared to the original long-delay \textit{ab initio} signal (blue \edit{dashed}) as well as a best-fit re-weighting of the adiabatic and nonadiabatic signal decomposition (red \edit{solid}). 
    }
    \label{fig:refit}
\end{figure}

Part of the ``enhancement, then decay'' behavior in this region originates from the initial n3s photoexcitation, which makes a transient positive contribution to the scattering above $Q=3.2$~$\mathrm{\AA}^{-1}$ in Fig.~\ref{fig:difference signals}(f). There is an additional contribution that is linked to the interplay between adiabatic and nonadiabatic dissociation. Figure~\ref{fig:pes and diagram}(b) shows that while the adiabatic channel reaches its final population within 250~fs, the nonadiabatic channel continues to gain population out to at least 600~fs. Meanwhile, Fig.~\ref{fig:refit}(a) plots the \textit{ab initio} scattering contributions of these two channels at pump-probe delays greater than 500~fs. \edit{N-D bond fission in the dominant nonadiabatic channel is accompanied by evolution of the electronic character of the excited state from n3s to n$\sigma^*$ \cite{ashfold__2010}. This evolution relocalizes electron density tightly around the ND$_2$ fragment,
pushing the region of enhancement to higher $Q$ in Fig.~\ref{fig:refit}(a) and somewhat outside the maximum $Q$ range of our detector. The adiabatic channel, on the other hand, undergoes a diabatic transition back to the nonbonding electron distribution that 
allows the valence electron to remain relatively delocalized at larger radius with respect to the nitrogen 1s electrons.
The distinct timescales of the two dissociation channels, as well as their distinct $Q$-dependent scattering, serve to drive the total integrated signal between $3.3$~\edit{$\mathrm{\AA}^{-1}$}$<Q<4.3$~$\mathrm{\AA}^{-1}$ lower as the reaction progresses.}


\edit{We used the computed scattering signals for adiabatic and nonadiabatic dissociation shown in Fig.~\ref{fig:refit}(a) as basis functions and computed the least-squares best fit to the long-delay experimental data (black points with errorbars in Fig.~\ref{fig:refit}(b)) with respect to the coefficients of each scattering signal.}
The resulting best-fit scattering signal (shown in solid red \edit{in Fig.~\ref{fig:refit}(b)}) yielded an \edit{experimentally determined} branching ratio of 6.3:1. The AIMS simulation (scattering signal shown in dashed blue \edit{in Fig.~\ref{fig:refit}(b)}) produced an expected nonadiabatic to adiabatic branching ratio of about 2.6:1 \cite{champenois_femtosecond_2023}, which is less consistent with the experimental data. \edit{Discussion of the apparent discrepancy between the branching ratio implied by the experimental data and the AIMS trajectory simulations is contained in the End Matter.}

In conclusion, we have performed a joint experimental-theoretical time-resolved X-ray scattering study of photoexcited ammonia that makes direct observation of the evolving electronic structure throughout the reaction with high time resolution. The prompt signature of the Rydberg valence electron is apparent when comparing the data to the \textit{ab initio} and independent atom model computed signals. The late-time data show the signatures of both adiabatic and nonadiabatic dissociation channels that can be distinguished via the electronic structure of the ND$_2$($\tilde X$) and ND$_2$($\tilde A$) fragments. The competition between these two channels 
\edit{produces} an observable 
\edit{reduction in high-$Q$ scattering} that can be related to \edit{their branching ratio.}
The work presented here temporally resolves the structural evolution of photoexcited ND$_3$ that had previously been captured in a single timebin in the prior MeV-UED study of the same system \cite{champenois_femtosecond_2023}. It also represents an important benchmark for future thoretical efforts.

We anticipate that near-term advances in X-ray techniques and pump-probe methodology at FEL facilities will enable the study of single-electron motion in a wide range of systems. The development of attosecond hard X-ray pulses \cite{huang_generating_2017,yan_terawatt-attosecond_2024} and sub-femtosecond timing schemes \cite{li_attosecond_2022} open the possibility of spatially tracking of electron wavepackets involved in charge migration \cite{worner_charge_2017}, strong-field ionization \cite{agostini_free-free_1979,freeman_above-threshold_1987,corkum_above-threshold_1989}, high-harmonic generation \cite{ferray_multiple-harmonic_1988}, and photoionization \cite{hentschel_attosecond_2001} for the first time. Improvements in the repetition rate and average brightness of FELs for enhanced data statistics and higher X-ray photon energies for enhanced spatial resolution will enable the extension of valence electron scattering studies to heavier molecules with more core electrons and more complex dynamics \cite{northey_extracting_2024}. Ultimately, the coupling of advanced TRXS experimental techniques with high-fidelity \textit{ab initio} modeling promises to enable a much richer understanding of electronic dynamics generally and their effects on photochemistry.

\begin{acknowledgments}
I.G. and A.M.G. were supported by the National Science Foundation. A.M.G. was additionally supported by an NSF Graduate Research Fellowship. P.H.B., J.P.C., P.L., D.G., E.G.C., and T.J.A.W. were supported by the AMOS program within the U.S. Department of Energy (DOE), Office of Science, Basic Energy Sciences, Chemical Sciences, Geosciences, and Biosciences Division. Use of the Linac Coherent Light Source (LCLS), SLAC National Accelerator Laboratory, is supported by the U.S. Department of Energy, Office of Basic Energy Sciences under Contract No. DE-AC02-76SF00515. This work was supported by the National Institutes of Health grant S10 OD025079. N.H.L. acknowledges start-up funding from the School of Engineering Sciences in Chemistry, Biotechnology and Health (CBH), KTH Royal Institute of Technology and the Swedish Research Council (Grant no.~2022-02871) as well as compute time from the National Academic Infrastructure for Supercomputing in Sweden (NAISS).
\end{acknowledgments}



%

\section*{End Matter on Experimental and Computational Details}

The TRXS experiment was performed at the Coherent X-ray Imaging (CXI) instrument \cite{boutet_coherent_2010} at the Linac Coherent Light Source (LCLS) \cite{emma_first_2010} free electron laser (FEL) using a gas-phase scattering configuration that has been described in prior publications \cite{stankus_advances_2020,ma_quantitative_2024-1}. The gas-phase ND$_3$ sample was photoexcited by a 200~nm, 80~fs FWHM duration UV pulse on the $\nu_2'=5$ peak in the photoabsorption vibrational progression as shown in Fig.~\ref{fig:pes and diagram}(c). The system was probed at variable pump-probe delays $\tau$ by a 10~keV ($\lambda=1.24$~\AA), $\sim35$~fs FWHM duration X-ray pulse generated by the LCLS. The pump and probe pulses were co-timed by an arrival time monitor, or ``time tool,'' which had a temporal resolution of better than 15~fs FWHM \cite{glownia_pumpprobe_2019}. The scattered X-rays were collected on a 4-megapixel Jungfrau detector at a distance of 83~mm from the scattering center, enabling momentum transfers $Q=\frac{4\pi}{\lambda}\sin\left(\theta/2\right)$ in the range between 0.37 and 4.40~\AA$^{-1}$ to be collected. Individual Jungfrau frames associated with each FEL shot were down-sampled by applying the appropriate centering and scattering corrections \cite{stankus_advances_2020,ma_quantitative_2024-1} and then azimuthally averaged in bins spaced linearly in $Q$. \edit{Azimuthally anisotropic scattering changes \cite{biasin_anisotropy_2018} were observed immediately following time zero on the order of 0.25\%, which then decayed away with the ensuing dissociation. Figure~S3 in the Supplemental Material \cite{noauthor_see_nodate} illustrates the anisotropic signal throughout the measured delay scan range.} Individual shots were then sorted into time bins and averaged appropriately (see the Supplemental Material \cite{noauthor_see_nodate} for details) to compute the time-resolved total scattering signal $S(Q,\tau)$.

The detailed dynamics of the photoexcited ND$_3$ system were modeled using \textit{ab initio} multiple spawning (AIMS) \cite{ben-nun_ab_2000} simulations that were reused from Ref.~\cite{champenois_femtosecond_2023}. The AIMS-computed time-resolved populations of the main reaction channels are shown in Fig.~\ref{fig:pes and diagram}(b). These trajectories were launched on the $\nu_2'=4$ transition \cite{cheng_absorption_2006}, one absorption peak lower in energy from the $\nu_2'=5$ transition pumped in the experiment shown in \edit{Fig.~S1 of the Supplemental Material \cite{noauthor_see_nodate}. This will primarily affect the branching ratio between adiabatic and nonadiabatic dissociation, as pumping a higher vibrational peak will make the adiabatic dissociation channel more readily accessible \cite{ma_first_2012}.}
As ND$_3$ dissociates along the N-D bond coordinate $R_\mathrm{ND}$ via the dominant nonadiabatic channel (orange solid trajectory, Fig.~\ref{fig:pes and diagram}(a)), the electronic character of the excited state evolves from n3s to n$\sigma^*$ \cite{ashfold__2010}. This evolution to n$\sigma^*$ character relocalizes electron density around the ND$_2$ fragment compared to the initially photoexcited n3s state, as can be seen in the electron amplitude isosurfaces in Fig.~\ref{fig:pes and diagram}(a). The adiabatic channel (green dashed trajectory, Fig.~\ref{fig:pes and diagram}(a)), meanwhile, retains its relatively delocalized lone pair electron distribution. Both the photoexcitation process and the nuclear geometry evolution across multiple potential energy surfaces are thus accompanied by significant rearrangement of the valence electronic structure, which can be followed by TRXS.

\section*{End Matter on the Dissociation Branching Ratio}

The apparent discrepancy between the branching ratio implied by the experimental data and the AIMS trajectory simulations must be explained in the context of the prior work on this system. Several previous studies \cite{biesner_state_1989-1,chatterley_timescales_2013,yu_tunneling_2014,zhu_representation_2012,ma_first_2012} have suggested that adiabatic dissociation can only occur for pathways that are both energetically accessible and avoid the conical intersection at the planar C$_{2v}$ geometry. Yarkony and coworkers concluded based on high-level theory that the onset for adiabatic dissociation of ND$_3$ should occur at the $\nu_2'=4$ absorption peak, and that the branching ratio should be about 4:1 at the $\nu_2'=5$ state pumped in the current experiment \cite{zhu_representation_2012,ma_first_2012}. Their results, while consistent with experimental measurements of the branching ratio for NH$_3$ \cite{biesner_state_1989-1}, overestimate the level of adiabatic dissociation for the $\nu_2'=6$ vibrational state of ND$_3$. 
Our experimental determination of the branching ratio at $\nu_2'=5$ supports the results of prior photofragment translational spectroscopy measurements \cite{biesner_state_1989-1}, suggesting that the adiabatic channel is significantly suppressed in ND$_3$ relative to that of NH$_3$ in a way that is not yet accounted for by the techniques in Ma \textit{et al.} \cite{ma_first_2012} and in our own AIMS simulations.

\end{document}